\documentclass {elsart}
\usepackage{graphicx}

\usepackage{graphics}
\usepackage{graphicx}
\usepackage{epsfig}

\usepackage{amssymb}
\usepackage{amsmath}

\begin{document}

\begin{frontmatter}



\title{Deformed Statistics Kullback-Leibler Divergence Minimization within a Scaled Bregman Framework}


\author[SRC]{R. C. Venkatesan\corauthref{cor}}
\corauth[cor]{Corresponding author.}
\ead{ravi@systemsresearchcorp.com}
\author[UNLP]{A. Plastino}
\ead{plastino@venus.fisica.unlp.edu.ar}

\address[SRC]{Systems Research Corporation,
Aundh, Pune 411007, India}
\address[UNLP]{IFLP-CCT, National University La Plata \&
National Research Council (CONICET)\\ C. C. 727,  1900-La Plata,
Argentina}

\begin{abstract}
The generalized Kullback-Leibler divergence (K-Ld) in Tsallis
statistics [constrained by the additive duality of generalized
statistics (dual generalized K-Ld)] is here reconciled with the
theory of Bregman divergences for expectations defined by normal
averages, within a measure-theoretic framework. Specifically, it
is demonstrated that the dual generalized K-Ld is a scaled Bregman
divergence. The Pythagorean theorem is derived from the minimum
discrimination information-principle using the dual generalized
K-Ld as the measure of uncertainty, with constraints defined by
normal averages.  The minimization of the dual generalized K-Ld,
with normal averages constraints, is shown to exhibit distinctly
unique features.

\end{abstract}

\end{frontmatter}

\section{Introduction}

The generalized (also, interchangeably, nonadditive, deformed, or
nonextensive) statistics of Tsallis' has recently been the focus
of much attention in statistical physics, complex systems, and
allied disciplines [1].  Nonadditive statistics suitably
generalizes the extensive, orthodox Boltzmann-Gibbs-Shannon
(B-G-S) one. The scope of Tsallis statistics has lately been
extended to studies of lossy data compression in communication
theory [2] and machine learning [3,4]. A critical allied concept
is that of relative entropy, also known as  Kullback-Leibler
divergence (K-Ld), which constitutes a fundamental
distance-measure in information theory [5]. The generalized K-Ld
[6] encountered in deformed statistics has been described by
Naudts [7] as a special form of f-divergences [8]. A related
notion is that of Bregman divergences [9]. These are
information-geometric tools of great significance in a variety of
disciplines ranging from lossy data compression and machine
learning [10] to statistical physics [11].

The generalized K-Ld in a Tsallis scenario (see Eq. (6) of this
Letter) is not a Bregman divergence, which constitutes a serious
shortcoming. This is unlike the case of the K-Ld in the B-G-S
framework, which is indeed a Bregman divergence [10]. This
forecloses the ability of the generalized K-Ld to extend to the case
of generalized statistics the bijection-property between exponential
families of distributions and the K-Ld, and other fundamental
properties of Bregman divergences, true in the B-G-S framework.
\textit{The consequence of the bijection property is that every
regular exponential family corresponds to a unique and distinct
Bregman divergence (one-to-one mapping), and, there exists a regular
exponential family corresponding to every choice of Bregman
divergence (onto mapping)}.  The bijection property has immense
utility in machine learning, feature extraction, and allied
disciplines [10, 12, 13].

A recent study [14] has established that the dual generalized K-Ld
is a scaled Bregman divergence in a discrete setting.  Further, Ref.
[14] has tacitly put forth the necessity of employing within the
framework of generalized statistics the dual generalized K-Ld (see
Eq. (7) of this Letter), a scaled Bregman divergence, as the measure
of uncertainty in analysis based on the minimum discrimination
information (minimum cross entropy) principle of Kullback [15] and
Kullback and Khairat [16]. \textit{Scaled Bregman divergences,
formally introduced by Stummer [17] and Stummer and Vajda [18],
unify separable Bregman divergences [9] and f-divergences [8]}.

At this juncture, introduction of some definitions  is in order.\\
\textbf{Definition 1} (Bregman divergences)[9]: Let $ \phi $ be a
real valued strictly convex function defined on the convex set $
\mathcal{S} \subseteq dom(\phi) $, the domain of $ \phi $ such that
$ \phi $ is differentiable on $ ri(\mathcal{S}) $, the relative
interior of $ \mathcal{S} $. The Bregman divergence $ B_\phi
:\mathcal{S} \times {\mathop{\rm ri}} \left( \mathcal{S} \right)
\mapsto [0,\infty) $ is defined as: $ B_\phi \left( {z_1 ,z_2 }
\right) = \phi \left( {z_1 } \right) - \phi \left( {z_2 } \right) -
\left\langle {z_1  - z_2 ,\nabla \phi \left( {z_2 } \right)}
\right\rangle $, where: $\nabla \phi \left( {z_2 } \right) $ is the
gradient of $ \phi $ evaluated at $
z_2 $. \footnote{Note that $\left\langle\bullet,\bullet\right\rangle$ denotes the inner product.  Calligraphic fonts denote sets.}\\

\textbf{Definition 2} (Notations)[18]: $\mathcal{M}$ denotes the
space of all finite measures on a measurable space
$(\mathcal{X},\mathcal{A})$ and $\mathcal{P}\subset \mathcal{M}$ the
subspace of all probability measures. Unless otherwise explicitly
stated \textit{P},\textit{R},\textit{M} are mutually
measure-theoretically equivalent measures on
$(\mathcal{X},\mathcal{A})$ dominated by a $\sigma$-finite measure
$\lambda$ on $(\mathcal{X},\mathcal{A})$. Then the densities defined
by the Radon-Nikodym derivatives
\begin{equation}
p = \frac{{dP}}{{d\lambda }},r = \frac{{dR}}{{d\lambda }},and,m =
\frac{{dM}}{{d\lambda }},
\end{equation}
have a common support which will be identified with $\mathcal{X}$.
Unless stated otherwise, it is assumed that $P,R\in \mathcal{P},
M\in \mathcal {M}$ and that $\phi:(0,\infty)\mapsto \mathcal{R}$ is
a continuous and convex function.

\textbf{Definition 3} (Scaled Bregman Divergences) [18] \textit{The
Bregman divergence }of probability measures \textit{P}, \textit{R}
\textit{scaled} by an arbitrary measure M on
$(\mathcal{X},\mathcal{A})$ measure-theoretically equivalent with
\textit{P}, \textit{R} is defined by
\begin{equation}
\begin{array}{l}
 B_\phi  \left( {P,R\left| M \right.} \right) \\
 = \int_\mathcal{X} {\left[ {\phi \left( {\frac{p}{m}} \right) - \phi \left( {\frac{r}{m}} \right) - \left( {\frac{p}{m} - \frac{r}{m}} \right)\nabla \phi \left( {\frac{r}{m}} \right)} \right]} dM \\
  = \int_\mathcal{X} {\left[ {m\phi \left( {\frac{p}{m}} \right) - m\phi \left( {\frac{r}{m}} \right) - \left( {p - r} \right)\nabla \phi \left( {\frac{r}{m}} \right)} \right]} d\lambda . \\
 \end{array}
\end{equation}
The convex $\phi$ may be interpreted as the generating function of
the divergence.

\textbf{Definition 4}[19, 20] :  Let $(\mathcal{X},\mathcal{A})$ be
a measurable space while symbols $P$,$R$ denote probability measures
on $(\mathcal{X},\mathcal{A})$. Let $p,r > 0$ denote
$\mathcal{A}$-measurable functions on the finite set $\mathcal{X}$.
A $\mathcal{A}$-measurable function $p : \mathcal{X} \mapsto
\mathcal{R}$ is said to be a probability density function (pdf) if
$\int_\mathcal{X}pd\lambda=1$. In this setting, the measure $P$ is
induced by $p$, i.e.,
\begin{equation}
P(E)=\int_E pd\lambda;\forall E \in \mathcal{A}.
\end{equation}
Definition 4 provides a principled theoretical basis to seamlessly
alternate between probability measures and pdf's as per the
convenience of the analysis.

 The generalized
K-Ld is defined in the continuous form as [7]
\begin{equation}
D_\phi  \left( {\left. p \right\|r} \right) =  - \frac{1}{\kappa
}\int_{\mathcal{X}} {p\phi \left( {\frac{r}{p}} \right)d\lambda }  =
- \frac{1}{\kappa }\int_{\mathcal{X}} {p\left[ {\left( {\frac{p}{r}}
\right)^\kappa   - 1} \right]d\lambda } ,
\end{equation}
where $p$ is an arbitrary pdf, $r$ is the reference pdf, and
$\kappa$ is some nonadditivity parameter satisfying: $ - 1 \le
\kappa  \le 1;\kappa  \ne 0 $. Here, (1) employs the definition of
the \textit{deduced logarithm }[7]
\begin{equation}
\omega _\phi \left( x \right) = \frac{1}{\kappa }\left( {1 - x^{ -
\kappa } } \right) \ .
\end{equation}
Specializing the above theory to the case of Tsallis scenario by
setting $\kappa=q-1$ yields the usual doubly convex generalized K-Ld
[6]
\begin{equation}
D_{K-L}^{q} \left( {\left. p \right\|r} \right) = \frac{1}{q-1
}\int_{\mathcal{X}} {p } \left[ {\left( {\frac{{p }}{{r }}}
\right)^{q-1}   - 1} \right]d\lambda.
\end{equation}
Note that the normalization condition is: $\int_{\mathcal{X}} {p
}d\lambda=1$.  This result is readily extended to the continuous
case.

The additive duality is a fundamental property in generalized
statistics [1].  One implication of the additive duality is that it
permits a deformed logarithm defined by a given nonadditivity
parameter (say, $q$) to be inferred from its \textit{dual deformed}
logarithm [1,7] parameterized by: $q^*=2-q$. Section 4 of this Letter
highlights an important feature of Tsallis measures of uncertainty
subjected to the additive duality when performing variational
minimization.

Re-parameterizing (6) by specifying: $q\rightarrow 2-q=q^*$ yields
the dual generalized K-Ld\footnote {Here "$ \rightarrow $" denotes a
re-parameterization of the nonadditivity parameter, and is not a
limit.}
\begin{equation}
\begin{array}{l}
D_{K-L}^{q^*}  \left( {\left. p \right\|r} \right) = \frac{1}{1-q^*
}\int_{\mathcal{X}} {p } \left[ {\left( {\frac{{p }}{{r }}}
\right)^{1-q^*}   - 1} \right]d\lambda\\
 =\int_{\mathcal{X}}p\ln_{q^*}\left( {\frac{{p }}{{r }}}
\right)d\lambda=\int_{\mathcal{X}}\ln_{q^*}\left( {\frac{{dP }}{{dR
}}}
\right)dP=D_{K-L}^{q^*}  \left( {\left. P \right\|R} \right).\\
\end{array}
\end{equation}

\textbf{Proposition 1}:  $D_{K-L}^{q^*}$ is jointly convex in the
pair $(p||q)$. Given probability mass functions $(p_q,q_1)$ and
$p_2,q_2)$, then
\begin{equation}
\begin{array}{l}
 D_{K - L}^{q^ *  } \left( {\left. {\lambda p_1  + \left( {1 - \lambda } \right)p_2 } \right\|\lambda q_1  + \left( {1 - \lambda } \right)q_2 } \right) \\
  \le \lambda D_{K - L}^{q^ *  } \left( {\left. {p_1 } \right\|q_1 } \right) + \left( {1 - \lambda } \right)D_{K - L}^{q^ *  } \left( {\left. {p_2 } \right\|q_2 } \right), \\
 \end{array}
\end{equation}
$\forall~\lambda\in[0,1]$.  This result seamlessly extends to the
continuous setting.

 An important issue to address concerns the manner in which expectation values are computed.
 Nonextensive statistics has employed a number of forms in which
expectations may be defined. Prominent among these are the linear
constraints originally employed by Tsallis [1] (also known as
\textit{normal averages}) of the form: $ \left\langle A
\right\rangle = \sum\limits_i {p_i } A_i $, the Curado-Tsallis (C-T)
constraints [21] of the form:  $ \left\langle A \right\rangle _q  =
\sum\limits_i {p_i^q } A_i  \ $, and the normalized
Tsallis-Mendes-Plastino (TMP) constraints [22] (also known as
\textit{$q$-averages}) of the form:  $ \left\langle {\left\langle A
\right\rangle } \right\rangle _q  = \sum\limits_i {\frac{{p_i^q
}}{{\sum\limits_i {p_i^q } }}A_i } \ $.\footnote{In this Letter,
$<\bullet>$ denotes an expectation.}  A fourth constraining
procedure is the optimal Lagrange multiplier (OLM) approach [23]. Of
these four methods to describe expectations, the most commonly
employed by Tsallis-practitioners is the TMP-one.

The originally employed normal averages constraints were abandoned
because of difficulty in evaluating the partition function, except
for very simple cases.  The C-T constraints were replaced by the TMP
constraints because: $\langle 1 \rangle_q \neq 1$. Recent works by
Abe [24] suggest that in generalized statistics expectations defined
in terms of normal averages, in contrast to those defined by
$q$-averages, are consistent with the generalized H-theorem and the
generalized \textit{Stosszahlansatz} (molecular chaos hypothesis).
Understandably, a re-formulation of the variational perturbation
approximations in nonextensive statistical physics followed [25],
via an application of $q$-deformed calculus [26].

The minimum K-Ld principle is of fundamental interest in information
theory and allied disciplines. The nonadditive Pythagorean theorem
and triangular equality have been studied previously by Dukkipati,
\textit{et. al.} [27,28]. These studies were however performed on
the basis of minimizing the generalized K-Ld using questionable
constraints defined by C-T expectations and $q$-averages.
\textit{The Pythagorean theorem is a fundamental relation in
information geometry whose form and properties are critically
dependant upon the measure of uncertainty employed, and, the manner
in which expectations (constraints) are defined}.

This Letter fundamentally differs from the studies in Refs. [27] and
[28] in a two-fold manner: $(i)$ the measure of uncertainty is the
dual generalized K-Ld (a scaled Bregman divergence), and $(ii)$ the
constraints employed are defined by normal average constraints,
whose use in generalized statistics has been revived by the
methodology of Ferri, Martinez, and Plastino [29].

At this stage, it is important to interpret the findings in Ref. [24] within the context of the equivalence relationships between normal averages, C-T, $q$-averages, and OLM forms of expectations derived in Ref. [29].   First, while Ref. [24] has suggested the inadequacy of $q$-averages on physics-based arguments, the equivalence relationships in [29] are purely mathematical in nature.  Next, [29] provides a mathematical framework to minimize Lagrangians using the Tsallis entropy employing normal averages expectations.

A notable consequence of minimizing the generalized K-Ld or the dual
generalized K-Ld using normal averages constraints is that the
expression for the posterior probability is
\textit{self-referential}[1].
 Specifically, the expression contains a function of the posterior probability,
 which is unknown and to be determined.  Fundamental differences in deriving the generalized
Pythagorean theorem in this Letter \textit{vis-\'{a}-vis} the
analysis presented in Refs.[27] and [28] lead to results which are
qualitatively distinct from both an information-geometric
 as well as a
statistical-physics perspectives.

Thus, this Letter establishes the Pythagorean decomposition of the
dual generalized K-Ld (a scaled Bregman divergence) within the
framework of deformed statistics for physically tenable normal
averages expectations. Such an analysis forms the basis to
generalize the analysis in [12] for information theoretic
co-clustering for mutual information based models. By definition,
co-clustering involves clustering of data that inhabits a $m\times
n$ matrix. Co-clustering has utility in a number of critical
applications such as text clustering [30], bio-informatics [31],
amongst others.

Note that for mutual information based models, defining the scaled
Bregman information as the normal averages expectation of the dual
generalized K-Ld [14], the Pythagorean theorem derived for the dual
generalized K-Ld in this Letter provides the foundation to extend the
optimality of minimum Bregman information principle [12], [32] which
has immense utility in machine learning and allied disciplines, and,
the Bregman projection theorem to the case of deformed statistics.
Finally, the Pythagorean theorem and the minimum dual generalized
K-Ld principle developed in this Letter serve as a basis to
generalize the concept of I-projections [33] to the case of deformed
statistics.

This Introductory Section concludes by establishing the qualitatively distinct nature of this Letter:    \begin{itemize}\item $(i)$\emph{This Letter generalizes and extends the analysis in Ref. [14]}. In Ref. [14], it was shown that the dual generalized K-Ld is a scaled Bregman divergence.  This was demonstrated in a discrete setting.    The generalization  is accomplished in Section 3 by demonstrating that this property also holds true in a continuous setting.  This is accomplished by expressing the Radon-Nikodym derivatives (1) as Lebesgue integrals (3).  Note that in a continuous measure-theoretic framework, the relationship ((1) and (3)) between probability densities and probability measures is transparent.  The extension of the generalization of the results derived in Ref. [14] is presented in Sections 4 and 5 of this Letter.\\

Section 4 takes advantage of the seamless relationship between probability densities and probability  measures in a continuous setting to perform minimization of the dual generalized K-Ld by employing (1) and (3).  First, the Lagrangian for the minimum dual generalized K-Ld defined by probability densities for normal averages expectations (17), which is characterized by Lebesgue integrals, is subjected to a straightforward transformation by invoking (1) and (3).  This step is followed by a simple minimization of the transformed Lagrangian with respect to the probability measure, which yields the minimum dual generalized K-Ld criterion (25) defined in terms of probability densities. \\

 This minimum dual generalized K-Ld criterion  is then employed as the basis to derive the Legendre transform relations (26).  The Legendre transform conditions, in conjunction with the Shore expectation matching condition [34],  are central in deriving the Pythagorean theorem for the dual generalized K-Ld with normal averages constraints (Eq. (40) in Section 5 of this Letter).  At this stage, it is necessary to explain the tenability of employing the Shore expectation matching condition in generalized statistics, given the finding in [24] that the Shore-Johnson Axioms [35] (notably Axiom III - system independence) are not applicable in generalized statistics which models complex systems whose elements have strong correlations with each other. In the Shore expectation matching condition (see Section 5 of this Letter), the correlations and interactions between elements are self-consistently incorporated into the probability density with which the expectation is evaluated. Specifically, the probability density is unambiguously determined during the process of minimizing the dual generalized K-Ld, using normal averages constraints.  Thus, the  Shore expectation matching condition is not adversely affected by the inapplicability of the Shore-Johnson Axioms when utilized in deformed statistics. \\

 \item $(ii)$  As stated above, the basis for establishing the dual generalized K-Ld as a scaled Bregman divergence, and, the subsequent derivation of the Pythagorean theorem for normal averages expectations is motivated by extending the theory of I-projections [33] to the case of generalized statistics, and the derivation of iterative numerical schemes (such as iterative scaling, alternating divergence minimization, and the EM algorithm) based on a candidate deformed statistics theory of I-projections [36].  For this, the candidate deformed statistics I-divergence between two probability densities $p$ and $q$ is to be \textit{strictly convex}. \\

     This is true for the case of the usual K-Ld, the generalized K-Ld, and as stated in Proposition 1 of this Section, also holds true for the dual generalized K-Ld.  In Ref. [7], a form of a generalized K-Ld which is Bregman divergences has been derived, and employed with normal averages constraints in (for example, Ba\u{gci, Arda, and Server} [37]).  However, it is convex only in terms of one variable and is unsuitable to the primary leitmotif of this study, i.e.  generalizing I-projections and the above stated iterative numerical schemes [33, 36] to the case of deformed statistics.  This form of the generalized K-Ld which is a Bregman divergence does appear to have applications in other disciplines, as demonstrated by Ref. [37], amongst other works.

 \end{itemize}

\section{Theoretical preliminaries}

The essential concepts around which this communication revolves are
reviewed in the following sub-sections.

\subsection{Tsallis entropy and the additive duality}
\ By definition, the Tsallis entropy, is defined in terms of
discrete variables as [1]
\begin{equation}
\begin{array}{l}
S_q \left( p \right) = -\frac{{1 - \int_\mathcal{X} {p^q }d\lambda }}{{1 - q}}; \int_\mathcal{X} {p}d\lambda  = 1. \\
\end{array}
\end{equation}
The constant $ q $ is referred to as the nonadditive parameter.
Here, (9) implies that extensive B-G-S statistics is recovered as $
q \to 1 $. Taking the limit $ q \to 1 $ in (9) and invoking
L'H\^{o}spital's rule, $ S_q \left( p\right) \to S\left( p \right) $,
i.e.,  the Shannon entropy.  Nonextensive statistics is intimately
related to \textit{q-deformed }algebra and calculus (see [26] and
the references within). The \textit{q-deformed} logarithm and
exponential are defined as [26]
\begin{equation}
\begin{array}{l}
\ln _q \left( x \right) = \frac{{x^{1 - q} - 1}}{{1 - q}}, \\
and, \\
\exp _q \left( x \right) = \left\{ \begin{array}{l}
 \left[ {1 + \left( {1 - q} \right)x} \right]^{\frac{1}{{1 - q}}} ;1 + \left( {1 - q} \right)x \ge 0 \\
 0;otherwise, \\
 \end{array} \right.
\end{array}
\end{equation}
In this respect, an important relation from \textit{q-deformed}
algebra is  the $q$-deformed difference [26]
\begin{equation}
\begin{array}{l}
\ominus _q x = \frac{{ - x}}{{1 + \left( {1 - q} \right)x}}\\
\Rightarrow \ln _q \left( {\frac{x}{y}} \right) = y^{q - 1} \left( {\ln _q x - \ln _q y} \right). \\
 \end{array}
\end{equation}
 The Tsallis entropy may be written as [1]
\begin{equation}
\begin{array}{l}
S_q \left( p \right) =  - \int_\mathcal{X} p ^q \ln _q p d\lambda. \\
\end{array}
\end{equation}

This Letter makes prominent use of the \textit{additive duality} in
nonextensive statistics. Setting $ q^*=2-q $, from (11) the
\textit{dual deformed} logarithm and exponential are defined as
\begin{equation}
\begin{array}{l}
 \ln _{q^*}  \left( x \right) =  - \ln _q  \left( {\frac{1}{x}} \right),\,\, and, \,\, \exp _{q^*}  \left( x \right) = \frac{1}{{\exp _q  \left( { - x} \right)}}. \\
 \end{array}
\end{equation}

The dual Tsallis entropy, and, the dual generalized K-Ld may thus be
written as
\begin{equation}
\begin{array}{l}
S_{q^*} \left( p\right) =  - \int_\mathcal{X} {p}  \ln _{q^*} p d\lambda, \\
and,\\
D_{K-L}^{q^*} \left[ p||r \right] = \int_{\mathcal{X}}{p} \ln _{q^*}
(\frac{{p}}{{r}})d\lambda,
\end{array}
\end{equation}
respectively. Note that the dual Tsallis entropy acquires a form
identical to the B-G-S entropies, with $ \ln_{q^*}(\bullet) $
replacing $ \log(\bullet) $ [2].

\section{Dual generalized K-Ld as a scaled Bregman divergence}
\textbf{Theorem 1}: Let $ t = \frac{z}{m},z = \frac{{dZ}}{{d\lambda
}},Z\in \mathcal{P}$, and, $m$ being the scaling. For the convex
generating function of the scaled Bregman divergence: $\phi(t)=t\,
ln_{q^*}t$, the scaled Bregman divergence acquires the form of the
dual generalized K-Ld: $B_\phi \left( {P,R\left| M=R \right.}
\right)=\int_{\mathcal{X}}{p} \ln _{q^*}\left(
\frac{{p}}{{r}}\right)d\lambda$.

\textbf{Proof}: \\
From (1) and (2)
\begin{equation}
\begin{array}{l}
B_\phi  \left( {P,R\left| M \right.} \right) \\
= \int_{\mathcal{X}} {\left[ {\frac{{p}}{{m }}\ln _{q^ *  } \frac{{p }}{{m }} - \frac{{r }}{{m }}\ln _{q^ *  } \frac{{r }}{{m }} - \left( {\frac{{p}}{{m}} - \frac{{r}}{{m }}} \right)\nabla {\frac{{r }}{{m}}} \ln_{q^*} \left( {\frac{{r }}{{m }}} \right)} \right]}dM  \\
= \int_{\mathcal{X}}{\left[ {p \ln _{q^ *  } \frac{{p }}{{m }} - p \ln _{q^ *  } \frac{{r }}{{m }} - \left( {p  - r } \right)\left( {\frac{{r }}{{m }}} \right)^{1 - q^ *  } } \right]}d\lambda  \\
\mathop  = \limits^{\left( a \right)} \int_X {\left\{ {pm^{q^ *   -
1} \left[ {\ln _{q^ *  } p - \ln _{q^ *  } r} \right] - \left( {p -
r} \right)\left( {\frac{r}{m}} \right)^{1 - q^ *  } }
\right\}d\lambda },
\end{array}
\end{equation}
where $(a)$ implies invoking the $q$-deformed difference (11) with
$q^*$ replacing $q$.  Setting $m=r$ in the integrand of (15) and
re-invoking (11) yields (7)
\begin{equation}
\begin{array}{l}
 B_\phi  \left( {P,R\left| M=R \right.}
\right)=\int_{\mathcal{X}}{p} \ln _{q^*}\left(
\frac{{p}}{{r}}\right)d\lambda =\int_{\mathcal{X}} \ln _{q^*}\left(
\frac{{dP}}{{dR}}\right)dP.
\end{array}
\end{equation}
This is a $q^*$-deformed f-divergence and is consistent with the
theory derived in Refs. [17] and [18], when extended to deformed
statistics in a continuous setting.
\section{Canonical distribution minimizing the Dual Generalized K-Ld }

Consider the Lagrangian
\begin{equation}
\begin{array}{l}
 L\left( {x,\alpha ,\beta } \right) = \int_{\mathcal{X}} {p\left( x \right)\ln _{q^ *  } \left( {\frac{{p\left( x \right)}}{{r\left( x \right)}}} \right)d\lambda \left( x \right)}  \\
  + \int_{\mathcal{X}} {\left( {\sum\limits_{m = 1}^M {\beta _m p\left( x \right)u_m \left( x \right)d\lambda \left( x \right) - \left\langle {u_m } \right\rangle } } \right)}  - \alpha \int_{\mathcal{X}} {\left( {p\left( x \right)d\lambda \left( x \right) - 1} \right)}  \\
 \mathop  = \limits^{\left( a \right)} \int_{\mathcal{X}} {\ln _{q^ *  } \left( {\frac{{p\left( x \right)}}{{r\left( x \right)}}} \right)dP\left( x \right)}    \\
  + \int_{\mathcal{X}} {\left( {\sum\limits_{m = 1}^M {\beta _m u_m \left( x \right)dP\left( x \right) - \left\langle {u_m } \right\rangle } } \right)}- \alpha \int_{\mathcal{X}} {\left( {dP\left( x \right) - 1} \right)} , \\
 \end{array}
\end{equation}
where $u_m, m=1,...,M$ are some $\mathcal{A}$-measurable
observables. In the second relation in (17), $(a)$ implies invoking
(3) and Definition 4.\footnote{Note that the second relation in (17)
utilizes the relation from (1), (3), and Definition 4 (after some
abuse of notation): $ p\left( x \right) = \frac{{dP}}{{d\lambda
}}\left( x \right) = \frac{{dP\left( x \right)}}{{d\lambda \left( x
\right)}}$.}  Here, the normal average expectations are defined as
\begin{equation}
\int_{\mathcal{X}}p(x)u_m(x)d\lambda(x)=<u_m>;m=1,...,M.\\
\end{equation}
The variational minimization with respect to the probability measure
$P$ acquires the form

\begin{equation}
\begin{array}{l}
 \frac{{\delta L\left( {x,\alpha ,\beta } \right)}}{{\delta P}} = 0 \\
  \Rightarrow \ln _{q^ *  } \left( {\frac{{p\left( x \right)}}{{r\left( x \right)}}} \right) + \sum\limits_{m = 1}^M {\beta _m u_m \left( x \right) - \alpha  = 0}.  \\
 \end{array}
\end{equation}
Thus
\begin{equation}
p\left( x \right) = r\left( x \right)\exp _{q^ *  } \left( {\alpha  - \sum\limits_{m = 1}^M {\beta _m u_m \left( x \right)} } \right). \\
 \end{equation}
 Thus, the posterior probability minimizing the dual generalized K-Ld is
\begin{equation}
\begin{array}{l}
 p\left( x \right) = \frac{{r(x)\exp _{q^ *  } \left( { - \sum\limits_{m = 1}^M {\tilde \beta _m^{q^ *  } \left( x \right)u_m \left( x \right)} } \right)}}{{\left( {1 + \left( {1 - q^ *  } \right)\alpha } \right)^{\frac{1}{{q^ *-1  }}} }}; \\
 \tilde \beta _m^{q^ *  } \left( x \right) = \frac{{\beta _m }}{{1 + \left( {1 - q^ *  } \right)\alpha }}. \\
 \end{array}
\end{equation}
\textit{Here, (21) highlights the operational advantage in employing
dual Tsallis measures of uncertainty, since they readily yield the
$q^*$-deformed exponential form as a consequence of variational
minimization when using normal average constraints.} Multiplying
(19) by $p(x)$, integrating with respect to the measure
$\lambda(x)$, and invoking (18) and the normalization condition:
$\int_{\mathcal{X}}p(x)d\lambda(x)=1$ yields
\begin{equation}
\int_{\mathcal{X}} {p\left( x \right)\ln _{q^ *  } \left(
{\frac{{p\left( x \right)}}{{r\left( x \right)}}} \right)d\lambda
\left( x \right)}  + \sum\limits_{m = 1}^M {\beta _m \left\langle
{u_m } \right\rangle  = \alpha }.
\end{equation}
From (21) and (22), the canonical partition function is
\begin{equation}
\begin{array}{l}
 \tilde Z\left( {x,\tilde\beta _m^{q^ *  } \left( x \right)} \right)
  = \left( {1 + \left( {1 - q^ *  } \right)\alpha } \right)^{\frac{1}{{q^ *   - 1}}} . \\
 \end{array}
\end{equation}
\textit{Note that $\tilde Z\left( {x,\tilde\beta _m^{q^ *  } \left(
x \right)} \right)$ and $\tilde\beta _m^{q^ *  }(x)$ are to be
evaluated $\forall x\in \mathcal{X}$}. This feature is exhibited by
the variational minimization of generalized K-Ld's and generalized
mutual informations employing normal average constraints [2,3]. From
(23)
\begin{equation}
\alpha  = \ln _{q^ *  } \left( {\frac{1}{{\tilde Z\left(
{x,\tilde\beta _m^{q^ *  } \left( x \right)} \right)}}} \right).
\end{equation}
From (21), (22), and (24), it is evident that the form of the
posterior probability minimizing the dual generalized K-Ld is
\textit{self-referential} [1].  Further, for: $ \left[ {1 - \left(
{1 - q^ *  } \right)\sum\limits_{m = 1}^M {\tilde\beta _m^{q^ *  }
\left( x \right)u_m } } \right] < 0 $, the canonical posterior
probability in (21): $p(x)=0$.  This is known as the \textit{Tsallis
cut-off condition} [1].  Substituting (24) into (22) yields the
minimum dual generalized K-Ld
\begin{equation}
D_{K - L}^{q^ *  } \left[ {p\left\| r \right.} \right] =   \ln _{q^
*  } \left( {\frac{1}{{\tilde Z\left( {x,\tilde\beta _m^{q^ *  } \left( x
\right)} \right)}}} \right) - \sum\limits_{m = 1}^M {\beta _m
\left\langle {u_m } \right\rangle }.
\end{equation}
From (25), the following Legendre transform relations are obtained
\begin{equation}
\begin{array}{l}
 \frac{{\partial D_{K - L}^{q^ *  } \left[ {p\left\| r \right.} \right]}}{{\partial \left\langle {u_m } \right\rangle }} =  - \beta _m,  \\
\frac{\partial }{{\partial \beta _m }}\ln _{q^ *  } \left(
{\frac{1}{{\tilde Z\left( {x,\tilde \beta _m^{q^ *  } \left( x
\right)} \right)}}} \right) =   \left\langle {u_m } \right\rangle . \\
 \end{array}
\end{equation}
\section{Pythagorean theorem for the dual generalized K-Ld}
\textbf{Theorem 2}:  Let $r(x)$ be the prior probability
distribution, and $p(x)$ be the posterior probability distribution
that minimizes the dual generalized K-Ld subject to a set of
constraints
\begin{equation}
\int_{\mathcal{X}} {p\left( x \right)u_m(x) d\lambda \left( x
\right)} = \left\langle {u_m } \right\rangle ;m = 1,...,M.
\end{equation}
Let $l(x)$ be any other (\textit{unknown}) distribution satisfying
the constraints
\begin{equation}
\int_{\mathcal{X}} {l\left( x \right)u_m(x) d\lambda \left( x
\right)} = \left\langle {w_m } \right\rangle ;m = 1,...,M.
\end{equation}
Then \\
$(i)$ $ D_{K - L}^{q^ *  } \left[ {l\left\| {p} \right.} \right] $
is minimum only if (Shore expectation matching condition)
\begin{equation}
\left\langle {u_m } \right\rangle  = \left\langle {w_m }
\right\rangle .
\end{equation}
$(ii)$ From (29)
\begin{equation}
\begin{array}{l}
 D_{K - L}^{q^ *  } \left[ {l\left\| {r} \right.} \right] = D_{K - L}^{q^ *  } \left[ {l\left\| {p} \right.} \right] + D_{K - L}^{q^ *  } \left[ {p\left\| {r} \right.} \right] \\
  + \left( {1 - q^ *  } \right)D_{K - L}^{q^ *  } \left[ {p\left\| {r} \right.} \right]D_{K - L}^{q^ *  } \left[ {l\left\| {p} \right.} \right]. \\
 \end{array}
\end{equation}

\textbf{Proof}:  Taking the difference between the dual generalized
K-Ld's yields
\begin{equation}
\begin{array}{l}
 D_{K - L}^{q^ *  } \left[ {l\left\| {r} \right.} \right] - D_{K - L}^{q^ *  } \left[ {l\left\| {p} \right.} \right] \\
  = \int_{\mathcal{X}} {l\left( x \right)\left[ {\ln _{q^ *  }
  \left( {\frac{{l\left( x \right)}}{{r\left( x \right)}}} \right) -
  \ln _{q^ *  } \left( {\frac{{l\left( x \right)}}{{p\left( x \right)}}} \right)} \right]} d\lambda \left( x \right), \\
 \end{array}
 \end{equation}
 while multiplying and dividing the integrand of (31) by $
\left[ {1 + \left( {1 - q^ *  } \right)\ln _{q^ *  } \left(
{\frac{{l\left( x \right)}}{{p\left( x \right)}}} \right)} \right] $
 leads to
 \begin{equation}
\begin{array}{l}
 D_{K - L}^{q^ *  } \left[ {l\left\| {r} \right.} \right] - D_{K - L}^{q^ *  } \left[ {l\left\| {p} \right.} \right] \\
 = \int_{\mathcal{X}} {l\left( x \right)\left\{ {\frac{{\left[ {\ln _{q^ *  } \left( {\frac{{l\left( x \right)}}{{r\left( x \right)}}} \right) - \ln _{q^ *  } \left( {\frac{{l\left( x \right)}}{{p\left( x \right)}}} \right)} \right]}}{{1 + \left( {1 - q^ *  } \right)\ln _{q^ *  } \left( {\frac{{l\left( x \right)}}{{p\left( x \right)}}} \right)}}} \right.} \left. {\left[ {1 + \left( {1 - q^ *  } \right)\ln _{q^ *  } \left( {\frac{{l\left( x \right)}}{{p\left( x \right)}}} \right)} \right]} \right\}d\lambda \left( x \right). \\
 \end{array}
 \end{equation}

Invoking now the definition of the $q^*$-deformed difference from
(11) (by replacing $q$ with $q^*$) results in: $ \frac{{\ln _{q^ * }
\left( {\frac{{l\left( x \right)}}{{r\left( x \right)}}} \right) -
\ln _{q^ *  } \left( {\frac{{l\left( x \right)}}{{p\left( x
\right)}}} \right)}}{{1 + \left( {1 - q^ *  } \right)\ln _{q^ *  }
\left( {\frac{{l\left( x \right)}}{{p\left( x \right)}}} \right)}} =
\ln _{q^ *  } \left( {\frac{{p\left( x \right)}}{{r\left( x
\right)}}} \right) $. Thus,  after re-arranging the terms (32)
results in
\begin{equation}
\begin{array}{l}
 D_{K - L}^{q^ *  } \left[ {l\left\| {p} \right.} \right] = D_{K - L}^{q^ *  } \left[ {l\left\| {r} \right.} \right] \\
 -\int_{\mathcal{X}} {l\left( x \right)\left\{ {\ln _{q^ *  } \left( {\frac{{p\left( x \right)}}{{r\left( x \right)}}} \right)\left[ {1 + \left( {1 - q^ *  } \right)\ln _{q^ *  } \left( {\frac{{l\left( x \right)}}{{p\left( x \right)}}} \right)} \right]} \right\}d\lambda \left( x \right)}  \\
=D_{K - L}^{q^ *  } \left[ {l\left\| {r} \right.} \right]
  - \int_{\mathcal{X}} {\left\{ {l\left( x \right)\ln _{q^ *  } \left( {\frac{{p\left( x \right)}}{{r\left( x \right)}}} \right)} \right.}  \\
 \left. { + \left( {1 - q^ *  } \right)\ln _{q^ *  } \left( {\frac{{p\left( x \right)}}{{r\left( x \right)}}} \right)D_{K - L}^{q^ *  } \left[ {l\left\| {p} \right.} \right]} \right\}d\lambda \left( x \right). \\
 \end{array}
 \end{equation}
 At this point we expand (33) and invoke (19), (24), and (28) to arrive at
\begin{equation}
\begin{array}{l}
 D_{K - L}^{q^ *  } \left[ {l\left\| {p} \right.} \right] \\
  = D_{K - L}^{q^ *  } \left[ {l\left\| {r} \right.} \right] - \ln _{q^ *  } \left( {\frac{1}{{\tilde Z\left(  \bullet  \right)}}} \right)\int_X {l\left( x \right)d\lambda \left( x \right)}  \\
  + \int_{\mathcal{X}} {l\left( x \right)\sum\limits_{m = 1}^M {\beta _m u_m(x) } d\lambda \left( x \right)}  \\
  - \left( {1 - q^ *  } \right)\ln _{q^ *  } \left( {\frac{{p\left( x \right)}}{{r\left( x \right)}}} \right)D_{K - L}^{q^ *  } \left[ {l\left\| {p} \right.} \right] \\
  = D_{K - L}^{q^ *  } \left[ {l\left\| {r} \right.} \right] - \ln _{q^ *  } \left( {\frac{1}{{\tilde Z\left(  \bullet  \right)}}} \right) + \sum\limits_{m = 1}^M {\beta _m \left\langle {w_m } \right\rangle }  \\
 - \left( {1 - q^ *  } \right)D_{K - L}^{q^ *  } \left[ {l\left\| p \right.} \right]\ln _{q^ *  } \left( {\frac{{p\left( x \right)}}{{r\left( x \right)}}} \right),
\end{array}
\end{equation}
where: $\int_{\mathcal{X}}l(x)d\lambda(x)=1$.  Note that: $\tilde
Z(\bullet)=\tilde Z(x,\tilde \beta_m^{q^*}(x))$. Multiplying and
dividing the fourth term on the RHS of (34) by $p(x)$ and
integrating over the measure $\lambda(x)$ yields
\begin{equation}
\begin{array}{l}
 D_{K - L}^{q^ *  } \left[ {l\left\| {p} \right.} \right] \\
 = D_{K - L}^{q^ *  } \left[ {l\left\| {r} \right.} \right] - \ln _{q^ *  } \left( {\frac{1}{{\tilde Z\left(  \bullet  \right)}}} \right) + \sum\limits_{m = 1}^M {\beta _m \left\langle {w_m } \right\rangle }  \\
 - \left( {1 - q^ *  } \right)D_{K - L}^{q^ *  } \left[ {l\left\| {p} \right.} \right]\frac{{\int_{\mathcal{X}} {p\left( x \right)\ln _{q^ *  } \left( {\frac{{p\left( x \right)}}{{r\left( x \right)}}} \right)d\lambda \left( x \right)} }}{{\int_{\mathcal{X}} {p\left( x \right)d\lambda \left( x \right)}
 }}.
\end{array}
\end{equation}
Now, setting $\int_{\mathcal{X}}p(x)d\lambda(x)=1$, (35) acquires
the form
\begin{equation}
\begin{array}{l}
 D_{K - L}^{q^ *  } \left[ {l\left\| {p} \right.} \right] \\
 = D_{K - L}^{q^ *  } \left[ {l\left\| {r} \right.} \right] - \ln _{q^ *  } \left( {\frac{1}{{\tilde Z\left(  \bullet  \right)}}} \right) + \sum\limits_{m = 1}^M {\beta _m \left\langle {w_m } \right\rangle }  \\
 - \left( {1 - q^ *  }
 \right)D_{K - L}^{q^ *  } \left[ {l\left\| {p} \right.} \right]D_{K - L}^{q^ *  } \left[ {p\left\| {r} \right.} \right],
\end{array}
\end{equation}
 and, with the aid of (25), (36) yields
\begin{equation}
\begin{array}{l}
 D_{K - L}^{q^ *  } \left[ {l\left\| {p} \right.} \right] = D_{K - L}^{q^ *  } \left[ {l\left\| {r} \right.} \right] \\
 - \ln _{q^ *  } \left( {\frac{1}{{\tilde Z\left(  \bullet  \right)}}} \right) + \sum\limits_{m = 1}^M {\beta _m \left\langle {w_m } \right\rangle }  \\
  - \left( {1 - q^ *  } \right)\left( {\ln _{q^ *  } \left( {\frac{1}{{\tilde Z\left(  \bullet  \right)}}} \right) - \sum\limits_{m = 1}^M {\beta _m \left\langle {u_m } \right\rangle } } \right)D_{K - L}^{q^ *  } \left[ {l\left\| {p} \right.} \right]. \\
 \end{array}
\end{equation}
 The minimum dual generalized K-Ld
condition is
\begin{equation}
\frac{{\partial D_{K - L}^{q^ *  } \left[ {l\left\| {p} \right.}
\right]}}{{\partial \beta _m }} = 0.
\end{equation}
This implies that the posterior pdf $p$ whose canonical form is
given by (21) not only minimizes: $D_{K - L}^{q^ *  } \left[
{l\left\| {p} \right.} \right]$, but also minimizes: $D_{K - L}^{q^
* } \left[ {p\left\| {r} \right.} \right]$ as well.  Subjecting (37) to (38)
and invoking the second Legendre transform relation in (26) yields
the Shore expectation matching condition [34] for the dual
generalized K-Ld
\begin{equation}
\left\langle {u_m } \right\rangle  = \left\langle {w_m }
\right\rangle .
\end{equation}
Substituting now (39) into (37) and invoking (25) allows one to
write
\begin{equation}
\begin{array}{l}
 D_{K - L}^{q^ *  } \left[ {l\left\| {p} \right.} \right] = D_{K - L}^{q^ *  } \left[ {l\left\| {r} \right.} \right] - D_{K - L}^{q^ *  } \left[ {p\left\| {r} \right.} \right] \\
  - \left( {1 - q^ *  } \right)D_{K - L}^{q^ *  } \left[ {p\left\| {r} \right.} \right]D_{K - L}^{q^ *  } \left[ {l\left\| {p} \right.} \right]. \\
 \end{array}
\end{equation}
The Pythagorean theorem for the dual generalized K-Ld with normal
average constraints has two distinct regimes, depending upon the
range of the dual nonadditive parameter
\begin{equation}
\begin{array}{l}
 D_{K - L}^{q^ *  } \left[ {l\left\| {r} \right.} \right] \ge D_{K - L}^{q^ *  } \left[ {l\left\| {p} \right.} \right] + D_{K - L}^{q^ *  } \left[ {p\left\| {r} \right.} \right];q^ *   > 1, \\
 D_{K - L}^{q^ *  } \left[ {l\left\| {r} \right.} \right] \le D_{K - L}^{q^ *  } \left[ {l\left\| {p} \right.} \right] + D_{K - L}^{q^ *  } \left[ {p\left\| {r} \right.} \right];0 < q^ *   < 1. \\
 \end{array}
\end{equation}
While Theorem 2 is called the Pythagorean theorem, (30) is referred
to as the nonadditive triangular equality for the dual generalized
K-Ld.  It is interesting to note that the expectation-matching
condition (29) has a form identical to the case of the B-G-S model
($q^*\rightarrow1$), and differs from that of the Pythagorean
theorem for the ``usual" form of the generalized K-Ld for the case
of constraints defined by C-T expectations and $q$-averages,
respectively [27, 28]. Also to be noted is the fact that the minimum
dual generalized K-Ld condition (38) is guaranteed. This differs
from the case of the $q$-averages constraints derived in previous
works [27, 28].  This feature is of importance when generalizing the
minimum Bregman information principle to the case of deformed
statistics.

\section{Summary and conclusions}
This Letter has proven that the dual generalized Kullback-Leibler
divergence (K-Ld) is a scaled Bregman divergence, within a
measure-theoretic framework.  Also, the Pythagorean theorem for the
dual generalized K-Ld has been established from normal average
constraints which are consistent with both the generalized H-theorem
and the generalized \textit{Stosszahlansatz} (molecular chaos
hypothesis) [24]. Qualitative distinctions of the present treatment
\textit{vis-\'{a}-vis} previous studies have been briefly discussed.

Ongoing work serves a two-fold objective: $(i)$ the Pythagorean
theorem for the dual generalized K-Ld derived herein has been
employed to provide a deformed statistics information geometric
description of Plefka's expansion in mean-field theory [38].  While details of this analysis are beyond the scope of this Letter, only a cursory overview of this analysis is presented herein.

Extending the procedure followed in [38] to obtain the mean-field equations [39], to the case of generalized statistics, a \textit{deformed statistics mean-field criterion} is obtained in terms of minimizing a dual generalized K-Ld.  This is accomplished by extrapolation of the information geometric arguments in [40] to the case of deformed statistics. Application of the Pythagorean theorem (40) results in a \textit{modified deformed statistics mean-field criterion}, which when subjected to a perturbation expansion employing results of the $q$-deformed variational perturbation theory
developed in [25], yields candidate deformed statistics mean-field equations; $(ii)$ the results obtained in this Letter serve
as the foundation to extend the sufficient dimensionality reduction
model [41] to the case of deformed statistics.  Results of these
studies will be published elsewhere.

\textbf{Acknowledgements}

RCV gratefully acknowledges support from \textit{RAND-MSR} contract
\textit{CSM-DI $ \ \& $ S-QIT-101155-03-2009}.  Gratitude is expressed to the anonymous reviewers for their invaluable inputs.


\begin{thebibliography}{99}

\bibitem{ex2}
C. Tsallis,
\newblock {\em Introduction to Nonextensive Statistical Mechanics: Approaching a Complex World,
Springer},
\newblock  Springer, Berlin, 2009.

\bibitem{ex2}
R. C. Venkatesan and A. Plastino,
\newblock { Physica A,}, \textbf{388} (2009) 2337.

\bibitem{ex2}
R. C. Venkatesan and A. Plastino,
\newblock{"Deformed statistics formulation of the information bottleneck
method"},
\newblock {\textit{Proceedings of the IEEE Int. Symp. on
Information Theory }}, 1323, IEEE Press, 2009.

\bibitem{ex2}
A. F. T. Martins, N. A. Smith, E. P. Xing, P. M. Q. Aguiar, and M.
A. T. Figueiredo ,
\newblock {J. Machine Learning Research (JMLR)}, \textbf{10} (2009) 935.


\bibitem{ex2}
T. Cover  and J. Thomas,
\newblock {\em Elements of Information Theory},
\newblock  John Wiley $  \&  $ Sons, New York, NY, 1991.

\bibitem{3}
 L. Borland, A. Plastino  and C. Tsallis,
\newblock {J. Math. Phys.} \textbf{39}  (1998) 6490.

\bibitem{3}
J. Naudts,
\newblock {Rev. Math. Phys.}, \textbf{16} (2004) 809.

\bibitem{ex2}
I. Csisz\'{a}r, Tud. Akad. Mat. Kutat{\o}Int. K\"{o}zl., \textbf{8}
(1963) 85.



\bibitem{3}
L. M. Bregman, USSR Comp. Math. Math. Phys., \textbf{7} (1967) 200.

\bibitem{3}
A. Banerjee, S. Merugu, I. Dhillon, and J. Ghosh,
\newblock {J. Machine Learning Research (JMLR)}, \textbf{6} (2005) 1705.

\bibitem{3}
S. Abe and G. B. Bag\c{c}i,
\newblock {Phys. Rev. E}, \textbf{71} (2005) 016139.

\bibitem{3}
A. Banerjee, I. Dhillon, J. Ghosh, S. Merugu, and ,D. Modha,
\newblock {J. Machine Learning Research (JMLR)}, \textbf{8} (2007) 1919.

\bibitem{3}
K. S. Azoury and M. K. Warmuth,
\newblock {Machine Learning }, \textbf{43} (2001) 211.


\bibitem{3}
R. C. Venkatesan and A. Plastino, Physica A, \textbf{390} (2011)
2749.

\bibitem{1}
S. Kullback, \textit{Information Theory and Statistics}, Wiley, New
York, 1959.

\bibitem{1}
S. Kullback and M. A. Khairat,  Ann. Math. Statist., \textbf{37}
(1966) 279.

\bibitem{4}
W. Stummer, Proc. Appl. Math. Mech., \textbf{7} (2007) 1050503.

\bibitem{4}
W. Stummer and I. Vajda, "On Bregman Distances and Divergences of
Probability Measures", 2009.  arXiv: 0911.2784.


\bibitem{ex2}
I. M. Gelfand, N. A. Kolmogorov, and A. M. Yaglom, Dokl. Akad. Nauk
USSR, \textbf{111} (1956) 745.



\bibitem{3}
M. S. Pinsker, \textit{Information and Information Stability of
Random Variables and Process}, Holden- Day, San Francisco, CA, 1960.

\bibitem{3}
E. M. F. Curado and C. Tsallis,
\newblock {J. Phys. A: Math Gen.} \textbf{24} (1991) L69.

\bibitem{3}
C. Tsallis, R. S. Mendes and A. R. Plastino,
\newblock {Physica A} \textbf{261} (1998) 534.

\bibitem{3}
S. Mart\'{i}nez, F. Nicol\'{a}s, F. Pennini  and A. Plastino,
\newblock {Physica A}, \textbf{286} (2000) 489.

\bibitem{3}
S. Abe,
\newblock {Phys. Rev. E}, \textbf{79}, (2009) 041116.

\bibitem{ex2}
R. C. Venkatesan and A. Plastino,
\newblock { Physica A,} \textbf{398} (2010) 1159.  Corrigendum [Physica A, \textbf{389} (2010) 2155].

\bibitem{4}
E. Borges,
\newblock {Physica A}, \textbf{340} (2004).




\bibitem{3}
A. Dukkipati, M. Narasimha Murty, and S. Bhatnagar, Physica A,
\textbf{361} (2006) 124.



\bibitem{c3}
A. Dukkipati, M. Narasimha Murty, and S. Bhatnagar, "Properties of
Kullback- Leibler cross-entropy minimization in nonextensive
framework",\textit{ Proceedings of IEEE International Symposium on
Information Theory(ISIT)}, 2374,  IEEE Press, 2005.

\bibitem{3}
G. L. Ferri, S. Martinez and A. Plastino,
\newblock {J. Stat. Mech.: Theory and Experiment} \textbf{2005(04)}
(2005) P04009.

\bibitem{3}
I. Dhillon, S. Mallela, and D. Modha, "Information-theoretic
co-clustering", \textit{Proceedings of the 9th International
Conference on Knowledge Discovery and Data Mining (KDD)}, 89, 2003.

\bibitem{3}
S. C. Madeira and A. L. Oliveira, IEEE Trans. Comput. Biology and
Bioinfo., \textbf{1}  (1) (2004) 24.

\bibitem{3}
I. Csisz\'{a}r, Annals of Statistics, \textbf{19} (1991) 2032.

\bibitem{3}
I. Csisz\'{a}r, Annals of Probability, \textbf{3}(1) (1975) 146.

\bibitem{3}
J. E. Shore, IEEE Trans. Inf. Th., \textbf{IT-27} (1981) 472.

\bibitem{3}
J. E. Shore and R. W. Johnson, IEEE Trans. Inf. Th., \textbf{IT-26} (1980) 26;
\textbf{IT-27} (1981) 472; \textbf{IT-29} (1983) 942.
\bibitem{3}
I. Csisz\'{a}r, P. C. Shields, Information Theory and Statistics: A Tutorial. Foundations and Trends in Communications and Information Theory, \textbf{1} 4 (2004) 417.

\bibitem{3}
G. B. Ba\u{g}ci, A. Arda, and R. Server, Int. J. Mod. Phys, \textbf{20} (2006) 2085.


\bibitem{3}
C. Bhattacharyya and S. Sathiya Keerthi, J. Phys.A: Math. Gen.,
\textbf{33} (2000) 1307.

\bibitem{3}

D. J. Thouless, P. W. Anderson, and R. G. Palmer, Phil. Mag.,\textbf{ 35} 3 (1977) 593.


\bibitem{3}

S. Amari, K. Kurata, and H. Nagaoka, IEEE Trsns. Neural Net.,\textbf{ 3} 2 (1992) 260.



\bibitem{3}
A. Globerson and N. Tishby,
\newblock {J. Machine Learning Research (JMLR)}, \textbf{3} (2003) 1307.





\end{thebibliography}
\end{document}